\documentstyle[11pt]{article}
\newcommand{\ds}{\displaystyle}

\begin{document}

\Large

\begin{center}
{\bf Free Particle Eigenfunctions  of Schrodinger Equation\\
 with Quantized Space-Time}
\end{center}

\vspace{.4in}

\begin{center}
Manjit Bhatia$^1$\\

\vspace{.2in}

Department of Mathematics, University of Maryland, College Park MD 20472, U.S.A. and Professor Emeritus, Bowie State University, Bowie MD U.S.A 20715\\
and\\
\vspace{.2in}
P. Narayana Swamy$^2$

\vspace{.2in}
Department of Physics, Southern Illinois University,
Edwardsville, IL 62026, U.S.A.
\end{center}

\vspace{.2in}

\normalsize

\begin{center}
\bf {Abstract}
\end{center}
It is well-known that the coordinate as a continuous variable, consisting of a set of all points between $0$ and $L$ contradicts the observability of measurement. In other words there might exist a fundamental length in nature, such as the Planck length $\lambda_P$, so that it is not possible to  measure a position coordinate with accuracy smaller than this fundamental length. It is therefore necessary  to investigate the formulation of quantum mechanics using only discrete variables as coordinates. To investigate all of quantum mechanics or any branch of physics from this approach is of course a daunting task and thus it is worthwhile to consider a specific simple problem in order to formulate the basic ideas. In this note we compare the solutions of Schrodinger equation for one-dimensional free particle under the usual space-time continuum with   those that are obtained when space-time is assumed to be quantized using a simple model. For this purpose, we replace the derivatives occurring in Schrodinger equation with the corresponding discrete derivatives. We compute the probability density (and the probability current) under the two scenarios; they turn out to be quite different in the two cases. We also obtain the operator identity for the commutator $[p,x]^q$ under the assumption of quantized space-time and contrast it with the usual commutator $[p,x]$.

\vspace{.2in}

 \noindent ($1$)  Electronic address: mbhatia@math.umd.edu\\

 ($2$) Electronic address: pswamy@siue.edu
 \vspace{.2in}

Keywords: { Discrete variable, non-continuous variable, discrete coordinates, discrete derivative, discrete integrals}\\

\vspace{.4in}

\begin{center}\large
{1.   Introduction}
\end{center}
\normalsize
\vspace{.4in}

In elementary (non-relativistic) quantum mechanics \cite{Gasiorowicz}, starting from the Hamiltonian of a (one-dimensional) free particle, $H={\ds \frac{p^2}{2m}}$, the time-dependent Schrodinger equation is obtained by the prescription
$H \rightarrow  i \hbar {\ds \frac{\partial}{\partial t}}$ and $p \rightarrow  -i \hbar {\ds \frac{\partial}{\partial x}}$. The resulting partial differential equation, for the wave function $\psi(x,t)$,
\begin{eqnarray}
\frac{-\hbar^2}{2m} \frac{\partial^2 \psi(x,t))}{\partial x^2} =i \hbar\frac{\partial \psi(x,t))}{\partial t}  \label{SEqn}
\end{eqnarray}
is separable. With $\psi(x,t)=T(t)U(x)$, (\ref{SEqn}) splits into two (ordinary) differential equations in the usual manner.
\begin{eqnarray}
\frac{dT(t)}{dt}+i\omega T(t) =0  \label{SEqnt} \\
\frac{d^2U(x)}{d^2 x}+k^2 U(x)=0  \label{SEqnx}
\end{eqnarray}
where $\omega ={\ds \frac{E}{\hbar}}$ and $k={\ds \sqrt{\frac{2mE}{\hbar^2}}}$.
The general solutions of the equations (\ref{SEqnt}) and  (\ref{SEqnx}) are given by
\begin{eqnarray}
T_\omega(t)=T_0 e^{-i\omega t} \label{SolEqnt} \\
U_k(x)=A e^{ikx} + B e^{-ikx} \label{SolEqnx}
\end{eqnarray}
The general solution of (\ref{SEqn}), (the so called plane wave solution) of the time-dependent Schrodinger equation can be written as:
\begin{eqnarray}
 \psi(x,t)=A e^{\ds i(kx-wt)} + B e^{\ds -i(kx+wt)}  \label{plane1}
\end{eqnarray}
Focusing on the plane wave moving in the positive $x$-direction, we have
\begin{eqnarray}
\psi(x,t)=A e^{\ds i (k x-\omega t)} \label{plane1p}
\end{eqnarray}
\\The corresponding probability density is given by
\begin{equation}\label{cprobd}
  P(x,t) = \psi^*(x,t)\psi(x,t) = |A|^2
\end{equation}
 Thus the probability density is a constant, independent of $x$ and $t$.
\vspace{.2in}

We can also compute the flux (i.e. the probability current) for the plane wave moving in the positive direction as {\bf[2]}
\begin{eqnarray}
J(x,t) = & {\ds \frac{\hbar}{2im} \left( \psi^* \frac{\partial \psi}{\partial x} - \frac{\partial \psi^*}{\partial x} \psi \right)}
\label{deflux} \\
= &(\hbar k/m)|A|^2 \label{cflux}
\end{eqnarray}
The flux $J(x,t)= (\hbar k/m)|A|^2$  also turns out to be a constant.

\vspace{.2in}

Motivated in part, by the possible existence  of a fundamental length in nature, such as the Planck length $\lambda_P$ \cite{Baez}, in this note we wish to solve the above Schrodinger equation for a free particle over quantized  space-time domain, and investigate what it might yield. In particular, we would be interested in learning how the probability density and the flux might be different when the underlying $(x,t)$ domains are quantized.

\vspace{.4in}

\begin{center}\large

{2.   Quantized Space-time}
\end{center}
\normalsize
\vspace{.4in}

We use the word ``quantized'' as synonymous with discreteized. Thus our quantized $x$-domain would simply be the discreteized  real line given by $R_\lambda=\{j_x \lambda | j_x \in Z \}$, where $\lambda$ is an unspecified ``fundamental'' length, such as the Planck length \cite{Baez}-\cite{Adler}. The Planck length is defined as $\lambda_P = \sqrt{G\hbar /c^3}$ and has the approximate numerical value, $1.6 \times 10^{-35}$ meter.  We shall henceforward drop the subscript $P$ for convenience. For simplicity, we express the quantized time domain in terms of $\tau = \lambda/c$, $c$ being the velocity of light. Thus $R_\tau=\{j_t \tau | j_t \in Z \}$. Here $j_x$ and $j_t$ are arbitrary integers. Note that for any integer $j_x$, the quantity $j_x \lambda$ corresponds to a particular point on the real line $R$. Similarly $j_t \tau$ gives us a specific value for time $t$ for any integer $j_t$. The continuum limit would thus correspond to $j_x \rightarrow \infty $ while $j_x \lambda \rightarrow x$ would be finite. Similarly for $j_t$ and $j_t \tau$.

\vspace{.2in}

With $R_\lambda$ and $R_\tau$ , respectively, the domain for $x$ and $t$,  the differential equations (\ref{SEqnt}) and (\ref{SEqnx}) become {\it difference equations} \cite{Hildebrand}:
\begin{eqnarray}
\Delta T^q(j_t \tau)+i\omega \tau T^q(j_t \tau) =0  \label{DSEqnt} \\
\Delta^2 U^q(j_x \lambda)+ k^2 \lambda^2U^q(j_x \lambda)=0  \label{DSEqnx}
\end{eqnarray}
where the difference operator $\Delta$ is defined as usual \cite{Hildebrand} and the superscript $q$ indicates that we are working with quantized space-time.

\vspace{.2in}

Thus we have
\begin{eqnarray}
 \Delta T^q(j \tau)=T^q((j+1) \tau)-T^q(j \tau) \label{defdel} \\
 \Delta^2 U^q(j \lambda) = U^q((j+2) \lambda) -2U^q((j+1) \lambda)+ U^q(j \lambda) \label{defdel2}
\end{eqnarray}
Substituting for  $\Delta T^q(j \tau)$ and $\Delta^2 U^q(j \lambda)$ in (\ref{DSEqnt}) and (\ref{DSEqnx}), we get the following recursion relations:
\begin{eqnarray}
 T^q((j_t+1) \tau)- (1-i\omega \tau) T^q(j_t \tau) =0  \label{Rect} \\
 U^q((j_x + 2) \lambda)-2U((j_x + 1))+ (1+ k^2 \lambda^2)U^q(j_x \lambda)=0  \label{Recx}
\end{eqnarray}
The general solution of the above recursion relations can be obtained by standard methods \cite{Hildebrand} and can be displayed as follows:
\begin{eqnarray}
 T^q(j_t \tau) = & T(0)(1-i\omega \tau)^{j_t}  \label{SRect} \\
 U^q(j_x \lambda) = & A(1+ik\lambda)^{j_x} + B(1-ik\lambda)^{j_x}  \label{SRecx}
\end{eqnarray}
With $\psi^q(j_x \lambda, j_t \tau)= T^q(j_t \tau)U^q(j_x \lambda)$,
we may refer to the above eigenfunction $\psi^q(j_x \lambda, j_t \tau)$ as {\it a plane wave over quantized space-time} or more simply {\it qst}-plane wave.

\vspace{.2in}

Confining ourselves to  the {\it qst}-plane wave, moving in the positive $x$-direction, we have
\begin{eqnarray}
\psi^q(j_x \lambda, j_t \tau)= A(1 + i k \lambda)^{j_x} \; (1 - i \omega \tau)^{j_t} \label{psiq}
\end{eqnarray}
 With $\psi^q$ as above, the corresponding probability density is given by
\begin{eqnarray}
 P^q(j_x \lambda, j_t \tau)=\psi^{q*} \psi^{q}=|A|^2 \; (1 + k^2 \lambda^2)^{j_x} (1 + \omega^2 \tau^2)^{j_t} \label{qProbd}
\end{eqnarray}
Computing the flux of $\psi^q(j_x \lambda, j_t \tau)$, using (\ref{deflux}), (with partial derivative being replaced by corresponding partial-difference operators), we obtain
\begin{eqnarray}
 J^q(j_x \lambda, j_t \tau)=|A|^2 \; \frac{\hbar k}{m}(1 + k^2 \lambda^2)^{j_x} (1 + \omega^2 \tau^2)^{j_t} \label{qstflux}
\end{eqnarray}
Thus we note that the probability density for the plane wave over quantized space-time, $ P^q(j_x \lambda, j_t \tau)$ is {\it not uniform}; in fact it is a monotonically increasing function of both $j_x$ (i.e. $x$) and $j_t$ (i.e. $t$), with range $(0, \infty)$ for $ P^q$. Hence it becomes transparent that, for {\it any} $\lambda > 0$,  the functional behavior of $P^q(j_x \lambda, j_t \tau)$ is radically different from that of $P(x,t)$ in (\ref{cprobd}) and presents us with quite a novel picture of the `plane wave' when space-time is taken to be quantized. A similar conclusion can be reached by comparing the flux $J(x,t)$ as given in $(\ref{cflux})$ with $J^q(j_x \lambda, j_t \tau)$ as given in (\ref{qstflux}).

\vspace{.2in}

The question of orthogonality and normalization will be investigated in the future.

\vspace{.4in}

\begin{center}\large

{3.   The standard continuum limit: $\lambda \rightarrow 0 $.}
\end{center}
\normalsize

\vspace{.4in}

We wish to show that the limit of the eigenfunction for {\it qst-} plane wave moving in the positive $j_x-$direction, as $\lambda \rightarrow 0$ is the eigenfunction of the plane wave (\ref{plane1p}).
The limiting process that we employ is as follows: Given a real number $x$, $\lambda \rightarrow 0$ and $j_x \rightarrow \infty$, in such a way that $j_x \lambda \rightarrow x$. Similarly, for any time $t$, as $\tau \rightarrow 0$ and $j_t \rightarrow \infty$, $j_t \tau \rightarrow t$.
With this in mind, we proceed to obtain the limit with (\ref{psiq}). In the following $``\lim"$ stands for $``(\lim_{\lambda, \tau \rightarrow 0, j_x, j_t \rightarrow \infty, j_x\lambda \rightarrow x, j_t\tau \rightarrow t })"$
\begin{eqnarray}
&\lim \psi^q(j_x \lambda, j_t \tau) \nonumber
=\lim A(1 + i k \lambda)^{j_x} \; (1 - i \omega \tau)^{j_t} \nonumber \\
&=\lim A(1 + i k \lambda j_x/j_x)^{j_x} \; (1 - i \omega \tau j_t/j_t)^{j_t} \nonumber \\
&= A e^{i (k x-\omega t)}
\end{eqnarray}

Since in the limit $\psi^q(j_x, j_t) \rightarrow \psi(x, t)$,
the limiting values of probability density $P^q$ (\ref{qProbd}) and
the flux $J^q$ (\ref{qstflux}) are also the same as $P$ as in (\ref{cprobd}) and $J$ as in (\ref{cflux}), respectively. This fact can also be verified by examination of the corresponding expressions for ($P^q$ and $P$) and ($J^q$ and $J$).

\vspace{.4in}

\begin{center}\large
{4.   The [p,x]-Commutator}
\end{center}
\normalsize
\vspace{.4in}

We recall that, with the prescription, $p \rightarrow  -i \hbar {\ds \frac{\partial}{\partial x}}$, the commutator $[p,x]=(px-xp)$,
  acting on a wave function $\psi(x)$ becomes
  \begin{eqnarray}
\left[p,x \right]\psi(x) &=\left(px-xp \right)\psi(x)  \nonumber \\
&= (-i \hbar)\left(\frac{d}{dx}(x \psi(x))- x \frac{d}{dx} \psi(x)\right) \nonumber \\
&= (-i \hbar)\left(\psi(x) +x \frac{d}{dx}\psi(x)- x \frac{d}{dx} \psi(x)\right)  \nonumber  \\
&= (-i \hbar) \psi(x) \label{com}
\end{eqnarray}
$\Rightarrow$
\begin{eqnarray}
[p,x]\psi(x)= -i\hbar \psi(x)   \label{com2a}
\end{eqnarray}
\\We then obtain the usual operator identity: $[p,x]=-i\hbar$.
\\\linebreak
\\With the goal of obtaining {\it qst}-version of the commutator, we replace the derivative operator $\ds \frac{d}{dx}$ by the difference operator
$ (1/\lambda) \Delta $. Then the commutator $[p,x]^q=(px-xp)^q$,
  acting on a {\it qst}-wave function $\psi^q(j \lambda)$ becomes
  \begin{eqnarray}
\left[p,x \right]^q\psi^q(j \lambda)&= (-i \hbar)\left(((1/\lambda) \Delta(j\lambda \psi^q(j\lambda)) - j\lambda((1/\lambda) \Delta(\psi(j\lambda)) \right)\   \nonumber \\
&= (-i \hbar)\left(((j+1) \psi^q((j+1)\lambda)) - j \psi^q(j\lambda)) -j \psi^q((j+1)\lambda) +j \psi^q(j \lambda) \right)\   \nonumber \\
&= (-i \hbar) \psi^q((j+1)\lambda) \label{comq}
\end{eqnarray}
\\ Thus in the {\it qst}-case we get:
\begin{eqnarray}
\left[p,x \right]^q \psi^q(j\lambda)=(-i \hbar) \psi^q((j+1)\lambda) \label{comq2}
\end{eqnarray}
\\We introduce a ``shift operator'', $\sigma$ which acts on  $\psi^q(j\lambda)$ by shifting the argument of the wave function one unit to the right i.e
$\sigma \psi(j\lambda)= \psi^q((j+1)\lambda)$
Then we can restate (\ref{comq2}) as
\begin{eqnarray}
\left[p,x \right]\psi^q(j\lambda)=(-i \hbar) \sigma \psi^q(j\lambda) \label{comq3}
\end{eqnarray}
\\This enables us to express the {\it qst}-commutator (indicated by the superscrpt $q$) as an operator identity:
\begin{eqnarray}
\left[p,x \right]^q=(-i \hbar)  \sigma  \label{comq4}
\end{eqnarray}
Referring to the equation (\ref{psiq}), focusing on the $x$-varible only, we can deduce that $\psi^q((j+1)\lambda)=(1+i(p \lambda)/\hbar) \psi^q(j\lambda)$. (Note that $k=p/\hbar$.) Then using this we can rewrite (\ref{comq2}) as
\begin{eqnarray}
\left[p,x \right]^q \psi^q(j\lambda)=(-i \hbar) (1+i(p \lambda)/\hbar) \psi^q(j\lambda) \label{comq21}
\end{eqnarray}
which implies the operator identity:
\begin{eqnarray}
\left[p,x \right]^q &=(-i \hbar)(1+i(p \lambda)/\hbar) \label{comq22}
\end{eqnarray}
We note that the right hand side is dimensionally correct and the steps leading to (\ref{comq22}) are a direct consequence of  the discretization procedure. It is possible to introduce some dynamics into the system and obtain a  different dependence on $p$ on the right hand side of (\ref{comq22}), such as what would be obtained according to quantum gravity \cite {Adler}.

It would be interesting to investigate the uncertainty relations that follow from the above commutator (\ref{comq22}). Other modified commutators $[p,x]$ have been studied in the literature (see for example, \cite{Kempf et al}.

\vspace{.4in}

\begin{center}\large
{5.   Summary}
\end{center}
\normalsize
\vspace{.4in}

With a simple model of quantized space-time we have presented a solution of a Schrodinger equation for a free particle. We have introduced the notion of finite differences replacing continuous derivatives to formulate the Schrodinger equation in describing the free particle. Comparison of our {\it qst}-solution  in  (\ref{psiq}), with the usual plane wave solution (\ref{plane1p}), reveals significant differences for the probability density, $P(x,t)$ and the probability current $J(x,t))$ in the two cases. Whereas for plane wave function, $P(x,t)$ and $J(x,t))$  are  constant, for the
{\it qst}-plane wave,  $P^q(j_x \lambda,j_t \tau)$ and $J^q(j_x \lambda,j_t \tau)$ are both monotonically (exponentially) increasing functions (with range $(0, \infty)$), of integers $j_x$ and $j_t$ that are used to describe the quantized space time in our model. We have also shown in (\ref{comq4}) how the commutator operator $[x,p]$ is modified in our model of quantized space-time.
\\\indent In view of the above noted differences that illustrate the properties of a free particle solution of Schrodinger equation with $x, t$ domains as continuum versus $x, t$ domains as quantized, it would be interesting to compare the results for other problems of quantum mechanics under these two models for space-time. We are aware that it is a daunting task to formulate all of quantum mechanics in quantized space-time. But we should be able to solve some of the elementary problems in our model of quantized space-time. While we should not expect to be able to easily formulate a future quantum theory with these investigations, based on a simplistic model of quantized space-time, but it is expected that an understanding of these examples under a simple model would be helpful.

\clearpage

\begin{center}
\large {Acknowledgments}

\vspace{.2in}

\end{center}
\normalsize   {\it MB} would like to thank the faculty and staff of Department of Mathematics, University of Maryland, College Park, MD, for their help, support and encouragement.

\end{document}